\documentclass[10pt,jcp,aip,showpacs,longbibliography,superscriptaddress,twocolumn]{revtex4-1}
\usepackage{epsfig,amsmath,amssymb,graphicx,color,calc,epstopdf}
\usepackage{color}
\usepackage{inputenc,bm}
\usepackage{hyperref}
\usepackage[mathscr]{euscript}
\definecolor{gris}{gray}{0.8}
\definecolor{grisbleu}{rgb}{0.47,0.6,0.82}

\newcommand{\hide}[1]{&{\textcolor{gris}{{#1}}}\\}
\newcommand{\killinpaper}[1]{\textcolor{grisbleu}{{#1}}}
\renewcommand{\hide}[1]{}
\renewcommand{\killinpaper}[1]{}

  \let\g=\gamma \let\d=\delta
\let\e=\varepsilon   
\let\s=\sigma  \let\f=\varphi 
   
\let\D=\Delta   
   
   \let\io=\infty

\def\to{\rightarrow}  

\newcommand{\wh}{\widehat}


\usepackage{xr}

\begin{document}
\title{Perspective: Gardner Physics in Amorphous Solids and Beyond}
\author{Ludovic Berthier}
\affiliation{Laboratoire Charles Coulomb (L2C), Universit\'e de Montpellier, CNRS, Montpellier, France}
\author{Giulio Biroli}
\affiliation{Laboratoire de Physique de l'Ecole Normale Sup\'erieure, ENS, Universit\'e PSL, CNRS, Sorbonne Universit\'e, Universit\'e de Paris, Paris, France}
\author{Patrick Charbonneau}
\affiliation{Department of Chemistry, Duke University, Durham,
	North Carolina 27708, USA}
\affiliation{Department of Physics, Duke University, Durham,
	North Carolina 27708, USA}
\author{Eric I. Corwin}
\affiliation{Department of Physics and Materials Science Institute, University of Oregon, Eugene, Oregon 97403, USA}
\author{Silvio Franz}
\affiliation{LPTMS, UMR 8626, CNRS, Univ. Paris-Sud, Universit\'e Paris-Saclay, 91405 Orsay, France}
\affiliation{
Dipartimento di Fisica, Universit\`a La Sapienza, Piazzale Aldo Moro 5, I-00185 Roma, Italy.}
\author{Francesco Zamponi}
\affiliation{Laboratoire de Physique de l'Ecole Normale Sup\'erieure, ENS, Universit\'e PSL, CNRS, Sorbonne Universit\'e, Universit\'e de Paris, Paris, France}

\begin{abstract}
One of the most remarkable predictions to emerge out of the exact infinite-dimensional solution of the glass problem is the Gardner transition. Although this transition was first theoretically proposed a generation ago for certain mean-field spin glass models, its materials relevance was only realized when a systematic effort to relate glass formation and jamming was undertaken. A number of nontrivial physical signatures associated to the Gardner transition have since been considered in various areas, from models of structural glasses to constraint satisfaction problems. This Perspective surveys these recent advances and discusses the novel research opportunities that arise from them.\footnote{This Perspective offers a progress report for the Simons collaboration ``Cracking the Glass Problem", of which the authors are some of the principal investigators.
While the authors are the sole responsible for the content of this Perspective, the underlying research originates from a much larger group of researchers. Their contributions are cited throughout and recognized in the acknowledgments.}
\end{abstract}

\date{\today}

\maketitle

\section{Introduction}
As recently as five years ago, the Gardner transition was but an exotic feature of abstract models known only by a cabal of statistical mechanicians. The recent surge in interest follows from the realization that mean-field models of structural glasses, which are exact in the limit of infinite dimension, $d\rightarrow\infty$, can also undergo such a transition. A coordinated effort to better understand the putative materials properties of this transition in real(istic) physical systems has since ensued.  Because these advances have taken place fairly rapidly and over a range of subfields, it can be challenging to piece them together into a forward vision. This Perspective aims to contextualize the many theoretical, numerical and experimental opportunities that lie ahead. By contrast to a recent review of hard sphere glasses~\cite{CKPUZ17}, the scope of this Perspective is thus both narrower and broader. We here focus exclusively on Gardner physics, but do so for a richer variety of glass formers and from a broader variety of viewpoints. We also include an array of recent results that help identify upcoming research challenges.

\begin{figure}
    \centering
    \includegraphics[width=.8\columnwidth]{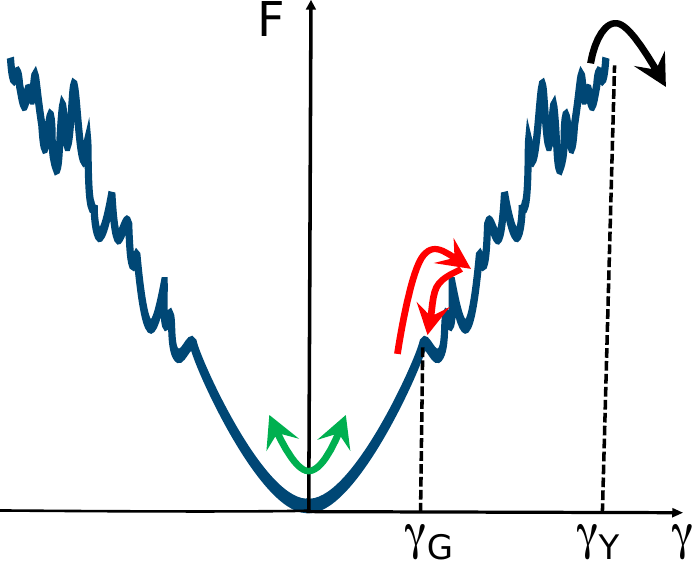}
    \caption{Schematic illustration of the evolution of a glass free energy basin under an applied perturbation $\g$, be it mechanical, thermal, or otherwise. At low $\g$, in the simple glass phase, the glass responds elastically and its free energy, $F$, is quadratic (green arrows). Beyond a Gardner transition $\g_{\rm G}$, the free energy basin becomes rugged and correspondingly the response becomes intermittent, due to jumps between sub-basins (red arrows). At even larger deformations, $\gamma_\mathrm{Y}$, the basin might disappear, or yield (black arrow). Adapted from Ref.~\onlinecite{JUZY18}.}
    \label{fig:freeenergybasin}
\end{figure}

To understand what a Gardner transition might look like in a materials context, we need to recall that a hallmark of solids is their elastic response to small deformations, i.e., their rigidity. In (perfect) crystals, rigidity is due to the spontaneous breaking of translational invariance that follows a first-order phase transition.
The resulting crystalline solid is \emph{thermodynamically} stable. In glasses and other amorphous solids, by contrast, rigidity emerges from the appearance of a multitude of long-lived metastable states. These solids also spontaneously break translational invariance, but only for a finite -- albeit long -- time. Rigidity is then a \emph{dynamical} phenomenon.
Once a system is confined to a metastable glass state, it responds elastically to an applied deformation, as would a crystal. Upon crossing the Gardner transition, however, that response changes spectacularly (Fig.~\ref{fig:freeenergybasin}). 
At the Gardner transition and in the Gardner phase that follows the materials response becomes strongly non-linear, and is characterized by intermittent `plastic events', all while the system remains globally solid. 
The breakdown of solidity happens at larger deformations, where the metastable glass state disappears entirely.

This Perspective elaborates on this point from a variety of outlooks. Section~\ref{sec:sg} revisits the original discovery of the Gardner transition in spin-glass models and presents its phase space description. Section~\ref{sec:hs} recapitulates its rediscovery in hard-sphere models, and Section~\ref{sec:ss} the limitations of its applicability in systems with softer pair interactions. Sections~\ref{sec:fr} and \ref{sec:xal} extend Gardner physics to the rheology of amorphous solids and to polydisperse crystals, respectively. Renormalization group (RG) considerations of the Gardner criticality are presented in Section~\ref{sec:rg}, and the connection between Gardner physics and constraint satisfaction problems is discussed in Section~\ref{sec:csp}. Section~\ref{sec:concl} briefly concludes.
 
\section{Gardner physics in spin glasses} 
\label{sec:sg}

In order to understand how the  Gardner transition emerges in the mean-field description of structural glasses, it is useful to revisit its original discovery in a class of mean-field spin glasses. We thus start this Perspective with a brief historical and conceptual detour through these abstract statistical mechanics models. 

In 1979, after years of theoretical struggle, the low temperature mean-field solution of the Sherrington-Kirkpatrick (SK) fully-connected spin glass model \cite{SK75} was  identified by Parisi \cite{P79}. This feat relied on the invention of the `replica symmetry breaking' (RSB) scheme. It took a few years to interpret physically~\cite{MPV87,CK94}, 
and decades to secure mathematically~\cite{Ta03,Pa13}, even just parts of the approach. As part of these efforts, in the 1980s several statistical physicists turned their attention to other fully-connected models amenable to the RSB scheme, such as Derrida's Random Energy Model (REM), and spin glasses with $p$-body \cite{D81}, Heisenberg \cite{GT81}, or Potts \cite{GKS85} interactions.  The REM family of models was especially important in elucidating the meaning of replicas, because these models could also be solved fairly straightforwardly using the Markov inequality %probabilistic methods available at the time
\cite{D85}. 

\begin{figure}[t]
    \centering
    \includegraphics[width=\columnwidth]{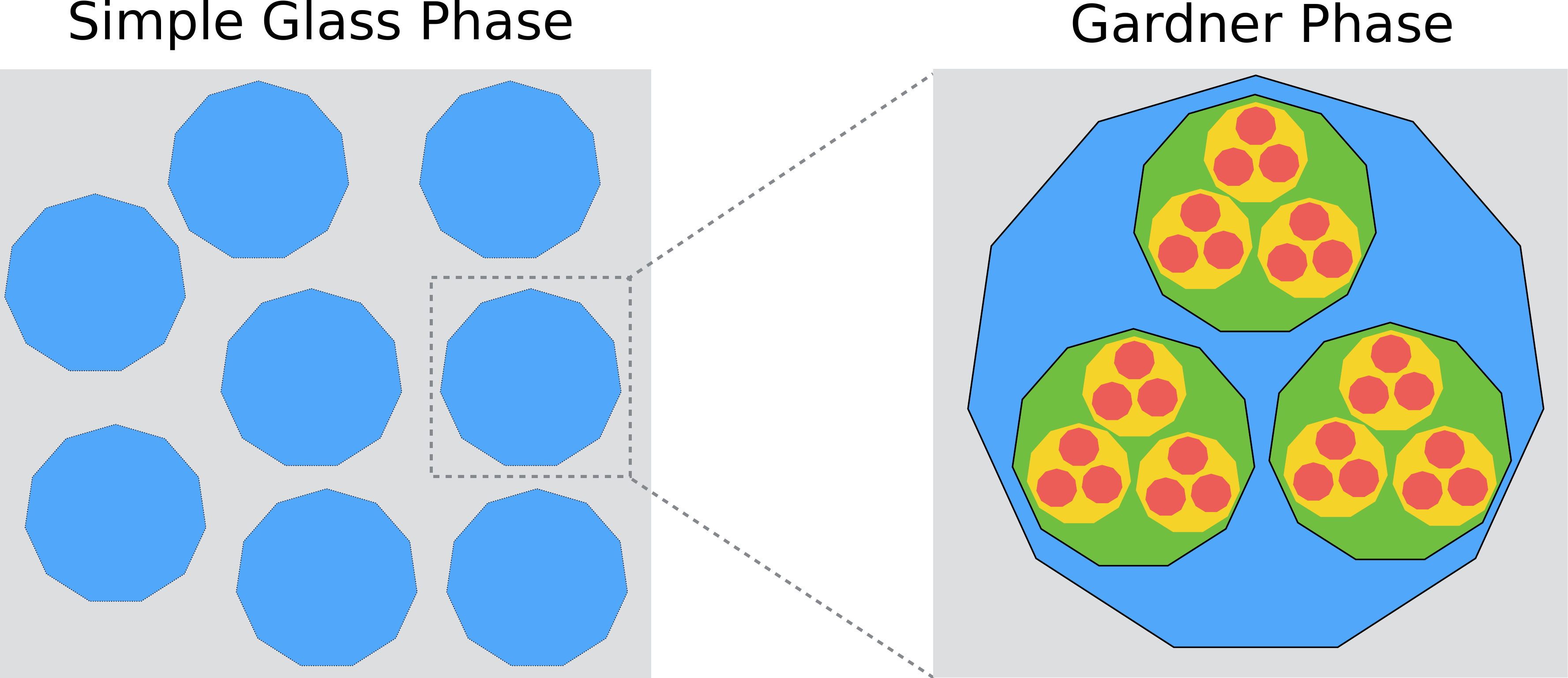}
    \caption{The phase space of a simple glass (REM-like) phase is formed of many distinct metastable clusters that are internally ergodic (blue regions), as in a perfect crystal, but separated by extensive barriers (grey regions). Upon crossing a Gardner transition the internal structure of these clusters changes dramatically. Each metastable glass state then fractures into an infinite hierarchy of sub-basins (green, yellow and red regions). Adapted from Ref.~\onlinecite{CKPUZ17}.}
    \label{fig:GardnerScheme}
\end{figure}

The decoding of RSB revealed its intimate relationship to ergodicity breaking. 
Spin glass models were then broadly sorted into two classes, depending of their type of ergodicity breaking and hence of RSB transition (see Ref.~\onlinecite{MPV87} for more details):
\begin{itemize}
  \item REM-like models, or discontinuous spin glasses, exhibit a thermodynamic transition of mixed first- and second-order character. Like at a first-order transition, the relevant (Edwards-Anderson overlap) order parameter jumps discontinuously 
and the associated susceptibility does not diverge, 
but like at a second-order transition no latent heat is released. The ergodic components that dominate the measure at low temperature are far in phase space from each other and are absolutely stable; different components have small Edwards-Anderson overlaps and are separated by extensive barriers. In the technical replica jargon, these models require a single step of RSB and are thus called 1RSB models (Fig.~\ref{fig:GardnerScheme}).
  \item SK-like or continuous spin glass models exhibit a thermodynamic second-order spin-glass transition that is accompanied with a divergent susceptibility. The whole spin glass phase is then marginal, ergodic components (or states) are critical and close to each other, and barriers between states are sub-extensive. Moreover, the organization of these states is both hierarchical and ultrametric, which corresponds to a maximally rich organization of the energy landscape. In the replica formalism, marginality is associated with the vanishing of the `replicon eigenvalue', which in technical jargon corresponds to `continuous' or full RSB (fullRSB).
\end{itemize}

In 1985, two papers further enriched this classification. Gross, Kantor and Sompolinsky, considering Potts glasses \cite{GKS85}, and Gardner, considering Ising $p$-spin glasses \cite{Ga85}, both noted that after a first phase transition to a 1RSB phase, these two REM-like models exhibit a fullRSB transition at a lower temperature (see Fig.~\ref{fig:GardnerScheme}). These reports of what was later called a `Gardner transition' were followed by others. Yet for most of the ensuing three decades these findings were only mentioned anecdotally, and treated as exotic features of exotic models. 

Meanwhile, the physical understanding of REM-like models markedly improved when Kirkpatrick, Thirumalai and Wolynes \cite{KTW89} recognized that their phenomenology is akin to that of structural glasses. They then realized that a first non-ergodicity, dynamical transition analogous to the transition in the mode-coupling theory of glasses~\cite{Go99,Go09}, at temperature $T_\mathrm{d}$, precedes the actual thermodynamic transition~\cite{KT87}. At $T_\mathrm{d}$ the phase space of these mean-field models becomes composed of an exponential (in the number of spins or particles) multiplicity of ergodic components that are dynamically inaccessible from one another. This multiplicity is measured by the complexity, which is akin to the entropy of a solid in excess of its vibrational contribution, i.e., its configurational entropy~\cite{BB11,doi:10.1063/1.5091961}. Beyond the dynamical transition, the complexity shrinks with temperature, and \emph{vanishes} at the thermodynamic transition at a finite temperature $T_\mathrm{K}$, as the configurational entropy seems to do in certain glass forming liquids, leading to an entropy crisis, as first noted by Kauzmann~\cite{Ka48}. 
On the basis of this analogy, the Random first order transition (RFOT) theory of REM-like models, which was understood to include structural glasses, was proposed~\cite{KTW89,LW07,WL12}. 

Over the following decades, the spherical $p$-spin model became the central focus of the statistical mechanics of disordered systems, as a prototypical
model of the RFOT universality class. 
Because this model does not exhibit a Gardner transition, the putative importance of this transition on structural glasses was left completely unexplored. Further consideration of Gardner transitions slowly reemerged in the late 1990s. Barrat, Franz and Parisi \cite{BFP97} showed them to be fairly ubiquitous (in essentially all spin glass models \emph{except} the spherical $p$-spin), and
Montanari and Ricci-Tersenghi \cite{MR03} revisited Gardner's work with more sophisticated theoretical tools--naming the transition by the same token. These studies were by then, however, quite far from the structural glass mainstream.

\section{Gardner physics in hard spheres}

\label{sec:hs}
\begin{figure}
    \centering
    \includegraphics[width=\columnwidth]{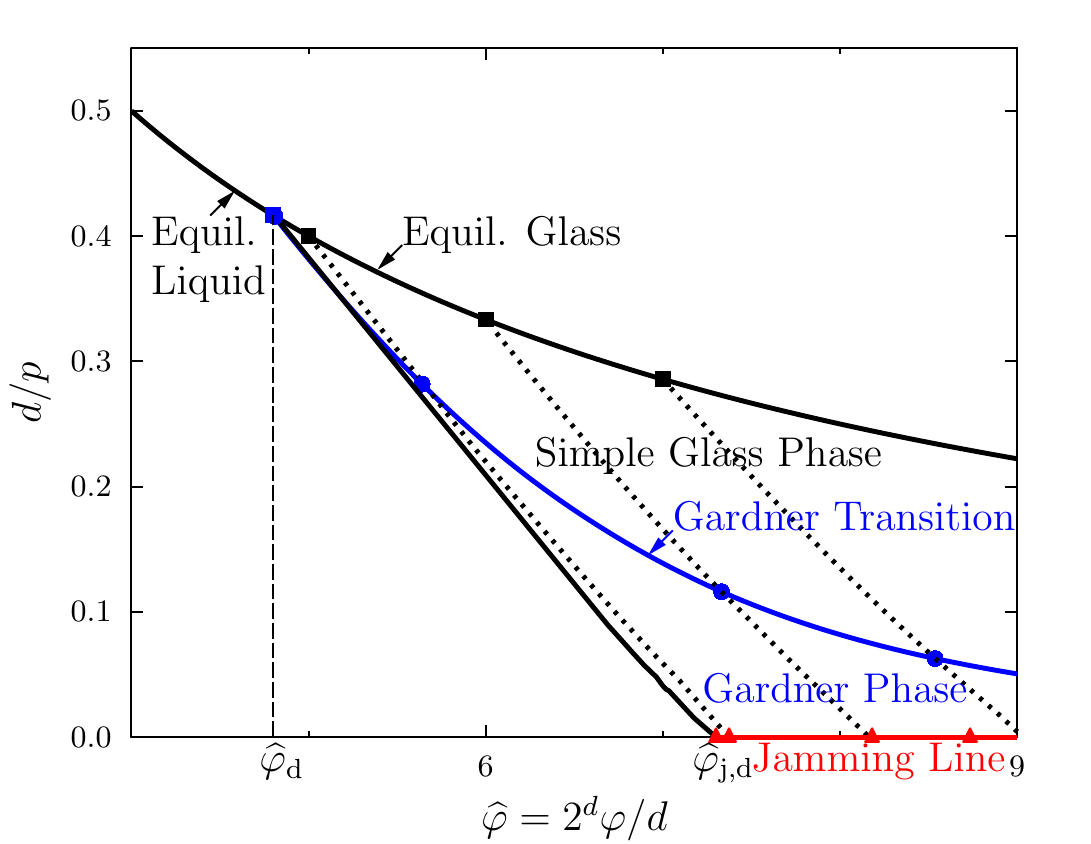}
    \caption{Phase diagram for hard spheres in the limit $d\rightarrow\infty$. At equilibrium (solid black line), for $\varphi<\varphi_\mathrm{d}$ the liquid is ergodic, while for $\varphi>\varphi_\mathrm{d}$ a large number of ergodic clusters are separated from one another, which gives rise to equilibrium glass states. Upon slowly compressing any such glass state (dashed lines), a Gardner transition (blue line) is encountered prior to reaching jamming (red line) at infinite (reduced) pressure, $p\rightarrow\infty$. Adapted from Ref.~\onlinecite{CKPUZ17}.}
    \label{fig:HSPD}
\end{figure}

Although the putative relevance of the Gardner transition to structural glasses was first suggested decades ago~\cite{KW87,KW87b}, 
only recently was Gardner physics fully brought to the field of amorphous solids. This advance resulted from the remarkable convergence of three concurrent yet independent lines of research, separately inspired by the seminal Liu-Nagel proposal for a unified description of amorphous solids~\cite{LN98}.
\begin{itemize}
\item Studies of dense sphere packings around jamming in two and three dimensions established through numerical simulations and analytical arguments that such packings are marginally stable~\cite{OLLN02,OSLN03,WNW05,WSNW05,BW06,BW09b,LN10,LNSW10,Wy12,LDW13,MW15}, in that they display an excess of low-energy modes that gives rise to anomalous scalings of physical quantities around jamming. 
\item Numerical studies of hard sphere fluids showed that glass formability increases with dimension~\cite{SDST06,vanmeel:2009b,CIMM10,CIPZ11}, 
which paved the way for numerical studies of the jamming transition in higher dimensions~\cite{SDST06,CIPZ11,CCPZ12,CCPZ15}.  
\item In the footsteps of Ref.~\onlinecite{KW87}, an analytical exploration of the glass phase of hard spheres in the mean-field limit $d\rightarrow\infty$  provided an analytical phase diagram within the simplest RS assumption~\cite{PZ05,PZ10,BJZ11a}.
\end{itemize}
Comparing these three sets of results gave rise to a conundrum. While the RS assumption
%assumed that jamming takes place within somewhat unremarkable ergodic clusters, and 
predicts \emph{hyperstatic} packings and simple critical exponents at jamming in $d\rightarrow\infty$~\cite{PZ10},
numerical simulations~\cite{SDST06,BW06,BW09b,CCPZ12} and analytical arguments~\cite{Wy12} showed that \emph{isostatic} packings and a non-trivial criticality are actually observed in all accessible~$d$. 

In order to solve this puzzle, an extended theoretical study of infinite-dimensional hard sphere glasses was undertaken~\cite{KPZ12}. The ensuing discovery of a Gardner transition~\cite{KPUZ13} and of the Gardner phase that lies beyond it, up to and including jamming~\cite{CKPUZ14,CKPUZ13}, put forward a physically coherent description of jamming.  The fullRSB solution associated with the Gardner phase indeed predicts that jammed sphere packings are isostatic and critically non-trivial in $d\rightarrow\infty$ as well. Building on this finding, the rest of the infinite-dimensional phase diagram of hard spheres was completed~\cite{RUYZ14,YZ14,RU16,CKPUZ17,footnote2} (see Fig.~\ref{fig:HSPD}). %, which remarkably predicts that all possible jamming points lie in the Gardner phase (see Fig.~\ref{fig:HSPD}). 
Additional scaling arguments~\cite{LDW13,DLBW14,DLW15} and analytical calculations~\cite{BU16} provided the critical scaling of the jammed phase and of finite-temperature systems.

In hindsight, the existence of a mean-field Gardner transition upon approaching jamming should not have been so surprising, because that same transition is found in essentially all REM-like models~\cite{GKS85,Ga85,BFP97,MR03,KZ08,KZ10,SCKLZ12,Ri13}, which includes infinite-dimensional hard spheres. (Hard spheres are nevertheless the first structural glass model without quenched disorder in which a Gardner transition was explicitly found~\cite{KPUZ13}.) 
What was rightfully surprising is the dimensional robustness of the jamming criticality, as discussed below. %~\cite{CKPUZ13,CKPUZ14}. 
In the rest of this section, we separately consider the Gardner transition, the Gardner phase, and jamming, and conclude by discussing possible paths towards experimental validations of the theory.

\subsection{Gardner Transition}
As further discussed in Sec.~\ref{sec:rg}, the very existence of a Gardner transition in $d<6$ was (and in some ways is still) far from obvious, based on renormalization group considerations. Hence, the first numerical study of this transition targeted a system from a family of idealized models in which particles interact in an abstract space with the topology of a random graph. This topology guarantees that long-range spatial fluctuations are fully suppressed and thus cannot extinguish mean-field critical points~\cite{MKK08}.
Note that in the mean-field phase diagram of Fig.~\ref{fig:HSPD}, the Gardner transition is very close to the dynamical transition at low densities~\cite{RUYZ14}. In that region, the system cannot easily be confined within a glass basin because of activated and hopping processes, and thus the Gardner transition cannot be properly detected.
Luckily enough, in the chosen Mari-Kurchan (MK) variant of these models (originally proposed by Kraichnan~\cite{Kr62}), equilibrium configurations can be generated at negligible computational cost~\cite{MK11,MPTZ11,CJPZ14} via the planting technique~\cite{KZ09}. This feature was essential to properly study the Gardner transition in this system~\cite{CJPRSZ15}.

With the MK results in hand as reference and with (contemporary yet unrelated) stunning improvements in sampling techniques that reach deep into the glass regime via the swap Monte Carlo algorithm~\cite{berthier2016equilibrium,NBC17}, the Gardner transition for hard sphere models in $d=2$ and $d=3$ could then be explored~\cite{BCJPSZ15}. This work first evinced that a growing correlation length, susceptibility, and correlation time qualitatively consistent with the mean-field Gardner transition can be observed in model glass formers. The numerical results obtained so far however only give a qualitative hint of a transition. %This is best appreciated by noting that the numerical signatures of a Gardner transition appear strong in dimension $d=2$~\cite{BCJPSZ15,LB18}, where it has been reasonably proposed that the Gardner transition cannot exist (see Sect.~\ref{sec:rg} for a discussion of the lower critical dimension of the Gardner transition).

A proper finite-size scaling study should more reliably test for the existence of a critical transition, but this project has not yet been undertaken. An extraction of the corresponding critical exponents is thus also missing. Such study would certainly be computationally challenging, but the relative robustness of the Gardner phenomenology in hard spheres compared to the equivalent transition in spin glasses~\cite{LPRR09,Janus12} (see Sec.~\ref{sec:rg}) suggests that the analysis might not be as painful as it first seems. Thanks to the recent extension of enhanced sampling techniques to higher dimensions~\cite{BCK19}, similar considerations in $d=4$ and 5 might also be within reach. The complex criticality scenarios proposed for this transition (see Sec.~\ref{sec:rg}) could thus be clarified. It may well be that the situations in dimensions $d=2$, 3 or 4 are all different, and future work should address this question.

\subsection{Gardner Phase and off-equilibrium dynamics}

The criticality of the Gardner phase beyond the transition is theoretically less controlled than the transition itself, and the corresponding numerical simulations are also less developed. Because equilibrating configurations in the Gardner phase requires prohibitively long simulation times,
one possibility is to study the off-equilibrium dynamics following quenches, as is routinely done in spin glasses~\cite{CK94,vincent1997slow}. Recent numerical works in this direction in $d=3$ in Ref.~\onlinecite{SZ18} and $d=2$ in Ref.~\onlinecite{LB18} are consistent with a complex organization of vibrational states separated by a hierarchy of barriers that gives rise to a rich aging behavior. 

Other than the complex aging behavior associated with a distribution of relaxation barriers, however, few insights have been clearly formulated. From a mean-field standpoint, solving the out-of-equilibrium dynamics in a fullRSB phase is a daunting challenge~\cite{CK94,MKZ15,AMZ19}. However, it should be possible to probe numerically and experimentally the state structure in that regime.

What has been robustly established is that after a direct quench from the simple glass phase past the estimated location of the Gardner transition, single-particle displacements exhibit a slow aging dynamics. Even after very large times, the system does not reach any kind of steady state. Dynamics inside a glassy basin must therefore proceed via a very slow exploration of a large number of states, separated by large barriers. Empirically, this implies that quenching two copies of the same hard sphere configuration to large pressures produces two distinct packings that are dynamically inaccessible from one another over the numerical time window. It has further been demonstrated that large correlation lengths develop and grow slowly with time after a quench~\cite{SZ18,LB18,De19}. The barriers that separate distinct packings thus correspond to highly collective and correlated particle displacements (as already suggested in Ref.~\onlinecite{BW09b}). Furthermore, by performing multiple  quenches from a given hard sphere glass configuration, it is found that a very large number of distinct packings can be produced, and that these different packings are organized hierarchically~%\cite{giorgiotobe,LB18}, 
\cite{LB18},
in a way that is strongly reminiscent of the fullRSB description. Alternatively, one can sample the nearby minima of the athermal jamming landscape and find that they are both hierarchical and ultrametric \cite{De19}. It is interesting to emphasize that the same phenomenology is found in $d=2$ hard disks, where no sharp Gardner transition is expected to take place. The phenomenology and complexity of the free energy landscape of hard spheres are therefore robust findings that should have direct experimental consequences, independently of the Gardner transition. 

\subsection{Jamming Criticality}

By contrast to the above two regimes, jamming criticality presents strong agreement between theory and numerics. The mean-field, $d\rightarrow\infty$ solution~\cite{CKPUZ13} and scaling arguments~\cite{Wy12} predict that at jamming the distribution of both small forces and small interparticle gaps exhibit non-trivial power-law scalings. Remarkably, numerical results for even $d=3$ and $d=2$ seemingly scale with those same non-trivial power-law exponents~\cite{SDST06,CCPZ12,LDW13,CCPZ15}, although without a clear theoretical basis for this superuniversality (see Sec.~\ref{sec:rg}).
The asymptotic scaling of the particle cage diameter upon approaching jamming~\cite{BW09b,CKPUZ14,DLBW14} 
and of the excitation spectrum~\cite{DLDLW14,FPUZ15,CCPPZ16,MSI17} match theoretical predictions similarly closely. This agreement is not only remarkable for the Gardner phase specifically, but is also the most notable materials prediction ever to emerge from a fullRSB analysis. Here again, no proper study of finite-size corrections yet validates these findings, but the robustness of existing numerical results here leaves little ambiguity.

\subsection{Experimental validation}
Experimental efforts to detect evidence of the Gardner transition are ongoing.  The first such evidence came in the form of a study of a quasi-thermalized two-dimensional hard-sphere--like granular system~\cite{SD16,Oh16}. However, more precise measurements in this vein may be difficult to achieve due to the need for systems to be well thermalized in order to clearly identify signatures of a Gardner transition.  The associated structural relaxation time is indeed well beyond experimental reach for the millimeter-size granular particles typically used to model glasses.  In this respect, colloidal systems offer a more fruitful avenue because it is possible to combine smaller-sized particles (allowing proper thermalization over experimental timescales) with light-scattering or microscopy techniques that are in principle able to resolve single particle dynamics even at very small length scales.   A second path to direct verification of a Gardner phase in physical systems is offered by the very short time dynamics of the glass.  Colloids within a stable glass experience a fixed cage structure and so the crossover from ballistic motion to caged behavior is expected to be simple and result in an exponential crossover in the mean squared displacement.  By contrast, in a Gardner phase cages are themselves ever shifting as the system explores the fractal free energy landscape.  Thus, colloids should be constantly leaving small cages only to find themselves in slightly larger cages \textit{ad infinitum}.  One thus expects a logarithmic growth of the mean squared displacement with lag time.  Intriguingly, a first study in this direction finds that glassy colloidal suspensions exhibit just such a transition from simple caged behavior to having logarithmically growing mean squared displacement~\cite{Ha19}.

The critical scaling of the mean squared displacement upon approaching jamming has also been measured in a colloidal glass\cite{ZDB17}, and the results
are here directly compatible with theoretical predictions.
Comparable validations of the force and gap criticality at jamming are, however, unlikely. 
The truly minute length and force scales over which power-law scalings are observed lie well below 
the resolution of any existing experimental setup. Indirect measures of the landscape structure are 
thus more likely to provide a real-world connection between amorphous solids and Gardner physics in the foreseeable future.

\section{Gardner physics in soft spheres}

\label{sec:ss}
Since the initial analysis of the glass phase of hard spheres, a much larger class of interaction potentials has been analysed within the realm of mean-field theory. A Gardner phase has been identified for a number of pair potentials and a broad range of physical conditions~\cite{BU16,BU18,SBZ18}. In short, the mean-field approach predicts that all types of pair potentials -- including soft repulsive spheres, Lennard-Jones and others -- can undergo a Gardner transition in some parts of their phase diagrams, as we now detail.  

\begin{figure}
\includegraphics[width=.45\textwidth]{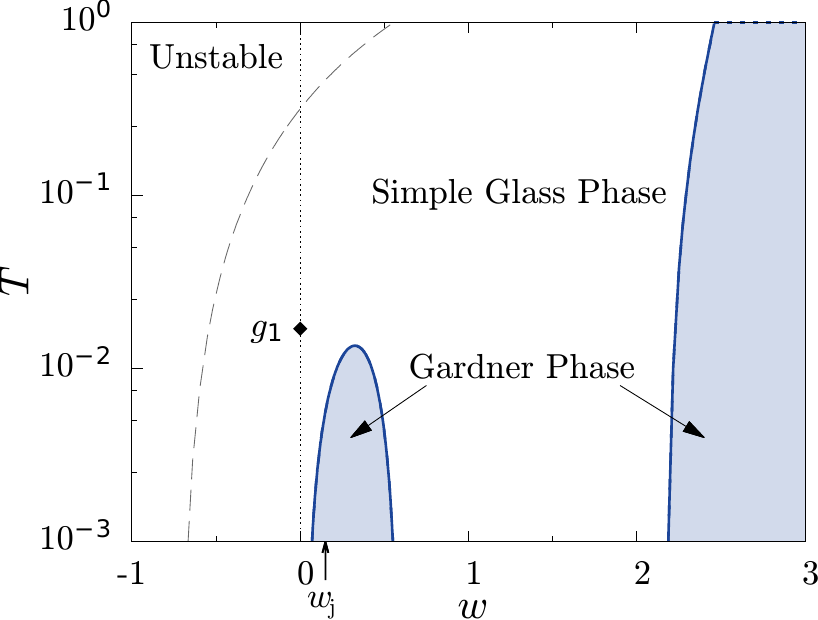}
\caption{
Mean-field phase diagram for soft repulsive WCA particles. The glass at point $g_1$ is prepared under equilibrium conditions at temperature $T_{g_1}$ and density $\rho_{g_1}$, and is then adiabatically followed up to different temperatures and densities, measured as $w =\ln [\rho / \rho_{g_1}]$. The glass state then either fluidizes (unstable region, beyond the dashed line), remains within an ergodic cluster (simple glass phase), or undergoes a Gardner transition (dark blue lines) to enter the Gardner phase (light blue areas). This example illustrates that in $d\rightarrow\infty$ soft glasses can generically undergo a Gardner transition under a variety of physical conditions, including around the jamming point $w_\mathrm{j}$ of the glass prepared at $g_1$. Adapted from Ref.~\onlinecite{SBZ18}.}
\label{fig:MFmarginal}
\end{figure}

A representative example (Fig.~\ref{fig:MFmarginal}) is the mean-field theoretical prediction for separately cooling and compressing an equilibrium glass state of soft repulsive spheres. The glass is of course unstable towards melting if the transformation takes it back to the fluid phase. 
More interestingly, it can undergo a Gardner transition to a marginally stable glass phase either by cooling or by compression. 
Generally, mean-field theory thus states that cooling a glass, or compressing a colloidal glass with repulsive interactions could lead to a marginally stable Gardner phase. 

The theoretical suggestion that all flavors of glass formers might undergo a Gardner transition initially raised the exciting possibility that the physical properties of a broad class of amorphous materials could be characterized by the low-lying collective excitations that characterise Gardner phases. This theoretical path could then potentially unify the physical properties of systems as diverse as granular materials near their jamming transition and atomistic glass-forming materials using the concept of marginal stability~\cite{MW15,KPUZ13,CKPUZ14}. 

A first step towards this quest aimed to confirm that a Gardner transition and the physics associated with the Gardner phase can be found in systems other than hard spheres. To this end, it is instructive to analyze models intermediate between hard spheres and standard Lennard-Jones glass formers, such as Weeks-Chandler-Andersen (WCA), harmonic or Hertzian particles. In these models of soft repulsive spheres, particles can overlap at a finite energetic cost, controlled by an energy scale $\epsilon$, if the interparticle distance is smaller than the interaction range, $\sigma$. The control parameters of the model are thus temperature and density, as in the mean-field analysis presented in Fig.~\ref{fig:MFmarginal}.  
In the limit of $\epsilon/k_\mathrm{B}T \to 0$ (which experimentally corresponds to relatively large colloids or stiff particles), these particles are still thermally agitated, but are also close to their jamming transition~\cite{BW09}. Computer simulations performed in this region of the phase diagram~\cite{SB19} confirm that all signatures of a Gardner transition (aging, slow dynamics, spatial correlations) can be observed in $d=3$, thus reflecting the emergence of a complex free energy landscape. These results show that the mean-field analysis remains qualitatively correct for three-dimensional models, at least in a region of the phase diagram close to the jamming transition.

Moving further away from this transition, computer simulations of Lennard-Jones particles in $d=2$ and soft repulsive spheres in $d=3$ have, however, revealed that a Gardner phase is not necessarily universally observed~\cite{SBZ17,SRDZ17}. 
Even when the transition is absent, the free energy basin of the simple glass phase breaks into several sub-basins at low temperatures~\cite{SRDZ17,SBZ17}, consistently with what has been observed in many numerical studies of the potential energy landscape of model glasses~\cite{Go69,SW82,Wales2003,He08}. A more refined analysis of the associated time and length scales, however, revealed that these sub-basins in fact correspond to a rare population of highly localized `defects'~\cite{SBZ17}. The physics of this phenomenon thus strongly contrasts with the mean-field prediction, which suggests that excitations should involve system-wide correlations. 
While this finding quashed hopes of universality, 
it also opened a way for a more systematic study of the interplay between extended defects, related to the mean-field Gardner phase, and localized defects, expected to be omnipresent in low-dimensional glasses~\cite{Go69,SW82,Wales2003,LW03,He08}. A promising line of research is the discovery that localized defects may themselves possess some universal properties~\cite{LDB16,PhysRevLett.121.055501,wang2019low}. The link between localized defects found in dense liquids far from jamming to collective excitations that proliferate close to jamming in marginally stable systems also remains to be understood~\cite{PhysRevE.98.060901}.   

Interestingly, the phenomenology of these localized defects is strongly reminiscent of the elementary excitations implicitly assumed by the empirical two-level system descriptions, which were first proposed more than 40 years ago but are still without a clear microscopic origin~\cite{AVW72,Ph87}.
The search for a Gardner phase in low-temperature glasses\cite{GLL18} may thus provide novel computational and experimental approaches to understand localized excitations in low-temperature glasses, and their interplay with more extended ones. Future work should aim at quantifying and generalizing these findings, to couple them with quantum mechanical descriptions that could explain the tunneling characteristics of two-level systems, and to quantitatively determine whether localized defects are indeed responsible for the cryogenic properties of amorphous solids. 

\section{Gardner physics in rheology}

Sections~\ref{sec:hs} and \ref{sec:ss} considered the emergence of a Gardner transition in a glass state that is either compressed or cooled, which are both isotropic transformations. Here, we  consider putative signatures of this same physics, but for the anisotropic transformations that emerge in the context of rheology. More specifically, we examine the stress-strain curves and the elastic response of glass states.

\subsection{Stress-strain curves}

\begin{figure}[t]
\includegraphics[width=.45\textwidth]{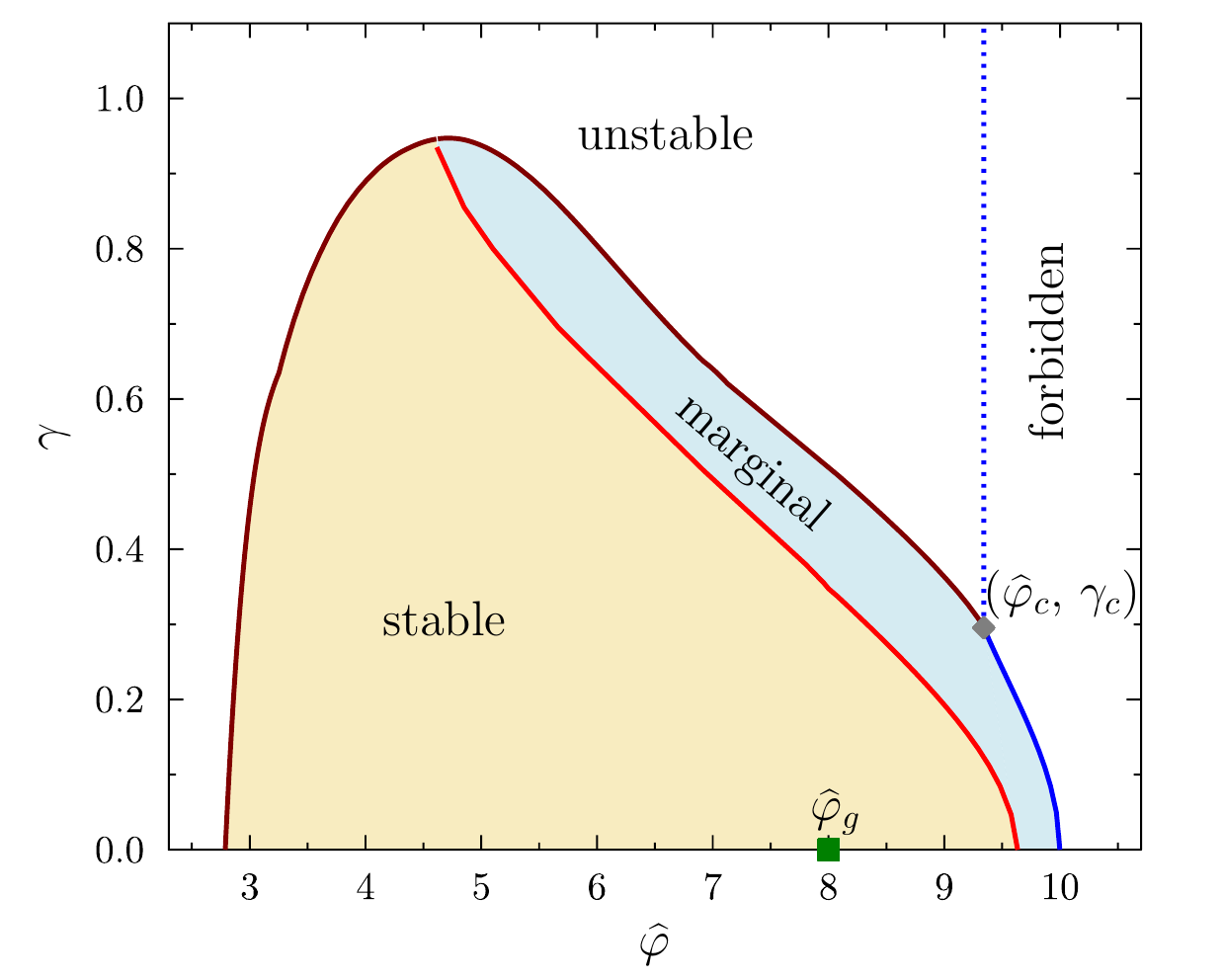}
\includegraphics[width=.45\textwidth]{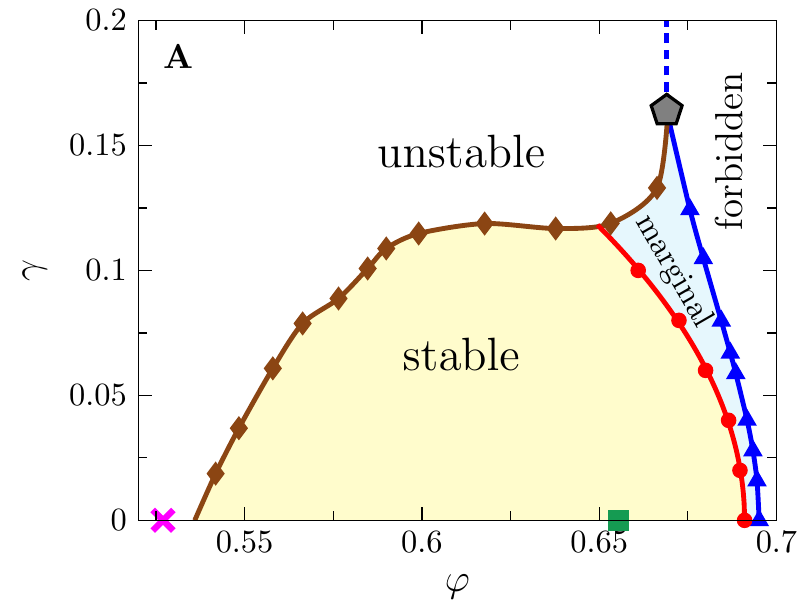}
\caption{
(Top) Theoretical shearing phase diagram of $d\rightarrow\infty$ hard spheres prepared in an equilibrium glass state at scaled packing fraction $\wh\f_g=8$, adapted from Ref.~\onlinecite{UZ17}.
The glass is first compressed at scaled packing fraction $\wh\f$ and then strained at shear strain $\g$, resulting in a $(\wh\f,\g)$ stability map. The simple glass (stable) phase is delimited by the yielding (brown) line at lower density, and by the shear jamming (blue) line at higher density. A Gardner transition (red) line delimits the Gardner (marginal) phase at high $(\wh\f, \g)$.
(Bottom) Corresponding numerical phase diagram of $d=3$ hard spheres prepared in equilibrium at $\f_g=0.655$, adapted from Ref.~\onlinecite{JUZY18}. The qualitative agreement between the mean-field prediction and the $d=3$ numerical simulations is remarkable, except in the vicinity of the critical point (grey diamond),
where the shear jamming and shear yielding lines merge. The qualitative difference might be explained either by the presence of
localized plastic excitations in $d=3$ systems (in which the role of the Gardner transition has yet to be clarified), or by the different nature of the yielding transition
in $d=3$ (shear band nucleation vs spinodal instability).
}
\label{fig:PDphigamma}
\end{figure}

The study of the rheology of amorphous solids by means of mean-field replica methods was initiated in Refs.~\onlinecite{YM10,Yo12},
and additional results for infinite-dimensional systems were obtained in Refs.~\onlinecite{RUYZ14,RU16,BU16,UZ17}. Motivated by these results, 
numerical studies in $d=3$ have considered elasticity, dilatancy, plasticity, marginal stability, yielding, and shear jamming~\cite{JY17,JUZY18}. 
For dynamically arrested equilibrium liquid configurations 
the initial response to shear is found to be perfectly elastic, and these materials can be characterized by a shear modulus $\mu = \mathrm{d} \s / \mathrm{d}\g$ 
and dilatancy $R = \mathrm{d} P / \mathrm{d} (\g^2)$, where pressure $P$ and shear stress $\s$ are conjugate observables (responses) to $\f$ and $\g$, respectively. Stress and pressure thus both increase under an applied strain at constant density. At larger strains, however, the response becomes
nonlinear and intermittent. In the mean-field description, the Gardner transition gives rise to this behavior, but in finite dimensions both collective and local excitations likely contribute~\cite{ML06,TTGB09,RTV11,ML11,HKLP11,ZS11,LLRW14,SLRR14,PVF16,HRC16,SLB19}. 
The glass then either yields or jams, as illustrated in the phase diagram of Fig.~\ref{fig:PDphigamma}. 
Yielding appears in $d\to\io$ as a spinodal point at which the glass solution is lost via a saddle-node 
bifurcation, leading to characteristic square-root singularities of physical observables~\cite{RUYZ14}, in particular $\s$ and $P$. 
In $d=3$, however, this singularity is smoothed by fluctuations, leading to a more complex behavior~\cite{PPRS17,JY17,JUZY18,OBBRT18}.
Jamming under shear corresponds to the joint divergence of $\s$ and $P$, and  its phenomenology is similar to its isotropic counterpart\cite{UZ17,JUZY18}.

\label{sec:fr}

\subsection{Linear response}

Glasses prepared beyond the Gardner transition display a markedly different elastic response. The simplest observable in this context is the shear modulus, $\mu$. In the stable solid phase, $\mu$ is given by a single number, i.e., the elasticity of the stable glass basin.
In the Gardner phase, by contrast, the $d\rightarrow\infty$ solution predicts that the linear response of the system takes a long time to relax~\cite{CKPUZ17}, because the ultrametric structure of
glass states is associated with an infinite hierarchy of time scales. 
After a short transient, the system equilibrates in the individual glass state in which it was prepared initially, 
the mean squared displacement approaches a constant value $\D_{\rm EA}$,
and correspondingly the stress decays, $\s(t) \sim \mu(\D_{\rm EA}) \g$. Waiting longer times (that diverge with the system size), 
allows the system to explore
a wider portion of the ultrametric tree of states. Correspondingly, the mean squared displacement increases, $\D>\D_{\rm EA}$,
and the stress further decays to $\s(t) \sim \mu(\D) \g$. The Gardner phase is then characterized by a function $\mu(\D)$ that gives the effective
shear modulus of the system~\cite{YZ14}, when it is allowed to explore phase space up to a mean squared displacement $\D$. The function
$\mu(\D)$ can be computed analytically in $d\to\io$, see Ref.~\onlinecite{YZ14}.
In particular, around jamming it is found that $\mu(\D_{\rm EA}) \sim P^\kappa$ while 
$\mu(\D < \D_{\rm EA}) \sim P \ll \mu(\D_{\rm EA})$, hence a dramatic softening should be observed if the system is allowed to leave the glass state in which it was prepared.
This theoretical prediction for $d\rightarrow\infty$ systems has been verified numerically~\cite{NYZ16,JY17} in $d=3$,
and might be related to the experimental findings of Refs.~\onlinecite{OH14,CSD14}.

\subsection{Nonlinear response}
Another key prediction of the $d\to\io$ solution concerns the nonlinear elastic response. Upon approaching the Gardner transition,
all the nonlinear elastic moduli are expected to diverge~\cite{BU16}. More precisely, consider the expansion of the elastic stress in powers of~$\g$,
\begin{equation}
\s = \mu_1 \g +  \frac1{3!} \mu_3 \g^3 + \cdots \ , \qquad \mu_n = \left.\frac{\mathrm{d}^n \s}{\mathrm{d}\g^n}\right|_{\g=0} \ ,
\end{equation}
where $\mu_1=\mu$ is the elastic shear modulus, and $\mu_3, \mu_5, \cdots$ are nonlinear response coefficients. These quantities are random variables because they depend on the particular glass sample under investigation.
Upon approaching the Gardner transition from
the stable phase, the average coefficient, $\mu_n$, remains finite, but sample-to-sample fluctuations diverge as $\d\mu_n^2 \sim \e^{3-2n}/V$, where $\e$ is the distance from the transition. 
Although fluctuations scale proportionally to $1/V$, the prefactor diverges for all $n> 2$. The linear response therefore remains well defined, but the sample-to-sample
fluctuations of the nonlinear moduli grow to be extremely large. 

The divergence of the fluctuations of elastic moduli is related to {\it plasticity}, i.e., the piece-wise linear behavior of the stress-strain curves in finite samples~\cite{HKLP11}.
A true divergence in thermodynamically equilibrated solids requires a divergent correlation length at the Gardner transition;
mean-field theory thus predicts plasticity to be extensive in the Gardner phase~\cite{BU16}.
An infinitesimal variation of density or shear strain in that phase indeed leads to extensive rearrangements, i.e., {\it avalanches}~\cite{LMW10,MW15,FS17}. The size distribution of these avalanches then scales as a power law with universal critical exponents, as has been confirmed
numerically~\cite{FS17}. 
In summary, while the response of the solid in the stable phase is elastic and reversible, in the Gardner marginally stable phase the response is plastic and irreversible,
leading to divergent nonlinear susceptibilities. 
It is important however to stress that in finite dimensions, localized plastic excitations involving a few particles are always present, independently of any Gardner transition~\cite{HKLP11};
these localized excitations can induce large fluctuations of the elastic moduli that should disappear in the thermodynamic limit, if equilibration within the glass basin can be reached. See Ref.~\onlinecite{JUZY18} for a more complete discussion of irreversibility and plasticity in this context.

Recent extension of the theory to the thermal regime using soft harmonic spheres\cite{BU18,SBZ18} will likely be followed by numerical work in finite dimension. An exciting possibility would be for these studies to provide a sharper understanding of the rheological role 
of localized excitations~\cite{ML06,TTGB09,RTV11,ML11,HKLP11,ZS11,LLRW14,SLRR14,PVF16,HRC16,SLB19}, 
as a counterpart of the comparable excitations observed in isotropically cooled or compressed systems.

\section{Gardner physics in Crystals}
\label{sec:xal}

\begin{figure}
\includegraphics[width=.45\textwidth]{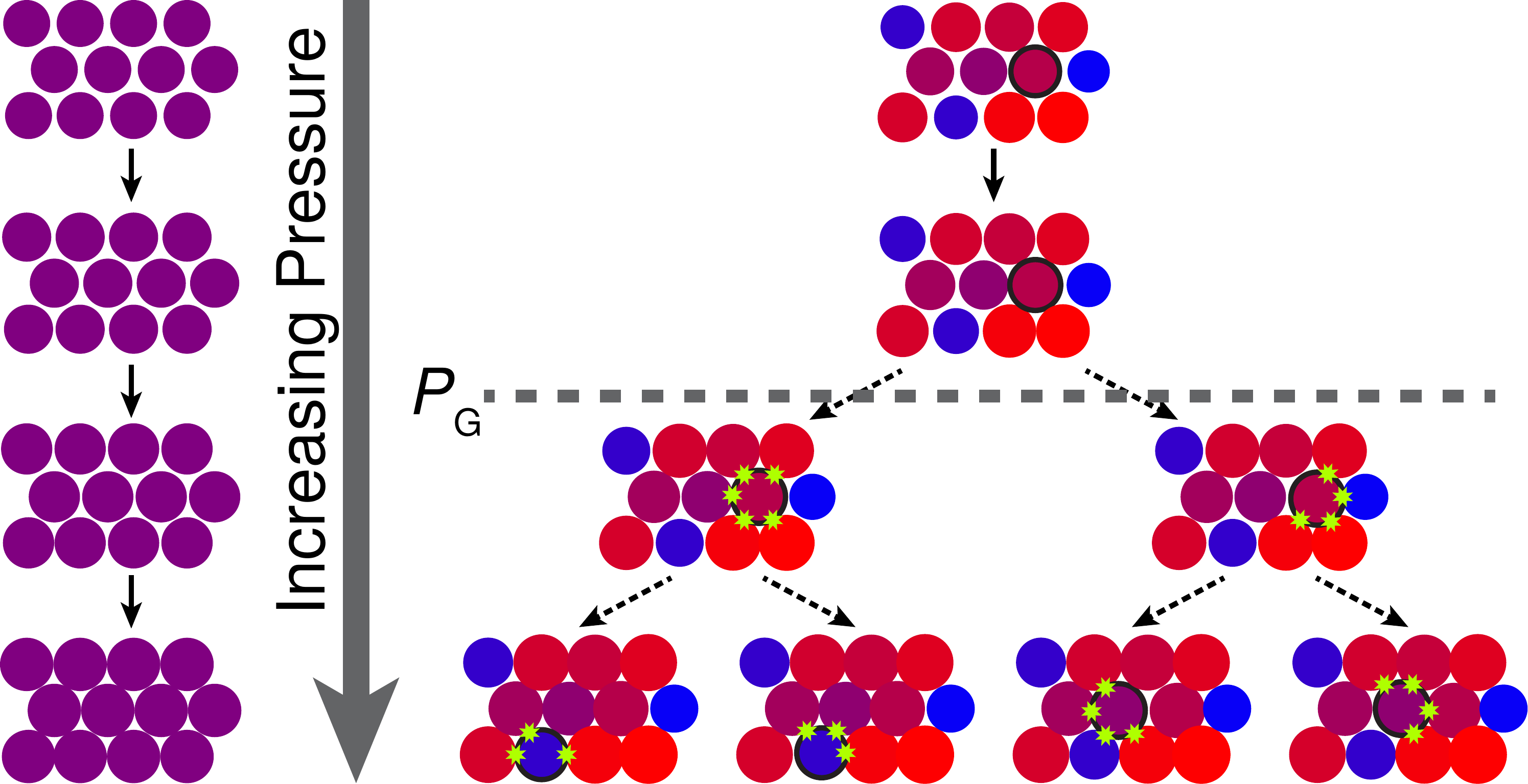}
\caption{Cartoon of a Gardner-like scenario for polydisperse crystals.  A perfect crystal (purple particles) has only one well-separated densest packing.  A crystal of polydisperse particles (colored from red to purple, according to their growing particle radius), by contrast, has a large number of nearly equivalent packings. Beyond the pressure onset of Gardner-like physics, at $P_\mathrm{G}$, such systems are forced into irreversible constraints. For the sake of illustration, consider the particle outlined in black. It is free to collide with all its nearest neighbors at low pressures, but forced to have one or another set of collisional contacts (green stars) as pressure increases. This overall process is cooperative, and thus cannot be reduced to such a simple local description, but this illustration can nonetheless be useful to understand and explain the physics at play. Adapted from Ref.~\onlinecite{CCFTN18}.}
\label{fig:CrystalGardnerIllustration}
\end{figure}

Although the Gardner transition has only been predicted for systems that are fully disordered, such as spin and structural glasses, the minimal conditions for its existence might not be so stringent. Several authors have indeed suggested that slightly disordered crystal packings display a marginal stability akin to that of fully disordered jammed packings~\cite{MKK08,GLN14,MZ16a}. In this section, we consider minimally disordered crystals that exhibit a behavior reminiscent of the Gardner physics.

It is straightforward to see how the disorder inherent to amorphous solids might give rise to a Gardner transition. The ultrametric structure of their landscape follows from the hierarchy of distinct vibrational states that emerge as the system is deformed (Fig~\ref{fig:CrystalGardnerIllustration}).  In perfectly crystalline materials, by contrast, every particle is contained within an identical and mechanically stable vibrational environment.  Crystals are thus marked both by highly degenerate overconstraints and the absence of any mechanism to break this degeneracy.  For instance, all particles in a face-centered cubic (FCC) crystal of hard spheres in $d=3$ have exactly $Z$ = 12 contacting neighbors at close packing, far in excess of the minimal isostatic criterion, $Z$ = 6, for marginal stability. An extensive number of contacts can thus be broken without destabilizing the system. However, this manifold degeneracy is easily destroyed by the introduction of even just a little dispersion in the particle size (polydispersity).  Such a perturbation explodes the high crystal symmetry into an enormous number of nearly identical, but distinct close-packed structures, each ultimately resulting from a unique set of correlated structural choices in the local environment.

Recent computer simulations of slightly polydisperse three-dimensional crystals\cite{CCFTN18} have identified Gardner-like and critical jamming signatures that are functionally identical to those found in fully amorphous systems in $d=3$: power-law scaling of forces near jamming, a growing relaxation time as the crystal basin breaks into smaller domains, followed by a rich aging behavior. 

The simplicity of these systems and of their underlying crystal order helps clarify the physical interpretation of the Gardner physics more generally. In particular, we learn from polydisperse crystals of hard particles that the position of the Gardner-like transition is controlled by the ratio of the spread of the polydispersity, $\delta$, to the typical interparticle spacing set by pressure, $h\sim1/P$. For $P\gg 1/\delta$, anomalous vibrational states are expected because the cage symmetry is then destroyed (Fig.~\ref{fig:CrystalGardnerIllustration}).  In amorphous solids, by contrast, the position of the  Gardner transition is known from mean-field theory to be controlled by the distance of the equilibrium glass state from the mean-field dynamical transition, but this relationship does not provide an intuitive microscopic mechanism for the transition.
The analogy with polydisperse crystals suggests that the Gardner transition emerges in amorphous solids from the geometric heterogeneity inherent to self-induced caging. Because this heterogeneity (as probed, for instance, by the long-time value of the dynamical susceptibility, $\chi_4$) monotonically decreases as the distance of the equilibrium glass state from the dynamical transition increases, we thereby obtain a more intuitive mechanism for the depression of the Gardner transition as the equilibrium glass density increases (see Fig.~\ref{fig:HSPD}).

The study of polydisperse crystals promises to further illuminate Gardner physics. From a theoretical standpoint, no explanation for their Gardner-like behavior exists beyond the intuitive analogies presented above. Developing a framework that could explain the similarities and differences between fully amorphous solids and  polydisperse crystals would likely provide additional insight into the microscopic processes at play. From a computational standpoint, the simplicity of polydisperse crystals makes them ideal for conducting finite-size scaling analysis of the Gardner criticality.  Such studies could prove determinative in assessing whether a thermodynamic phase transition or a crossover is actually at play. A systematic consideration of the jamming  criticality could similarly be undertaken. From an experimental standpoint, these systems offer great promise for validating Gardner physics in the laboratory. As we discussed in Sec.~\ref{sec:hs}, thermalizing colloidal or granular amorphous solids to the regime of interest to observe the Gardner transition is a very challenging experimental project. By contrast, crystals of hard grains can be manually prepared close to equilibrium, and colloidal crystallization of hard polydisperse particles is reasonably facile~\cite{PV86}. Because both families of materials are commercially available with a sufficiently low polydispersity (typically, $\delta\sim$ 1-10 \%) to distinguish crystal formation from the Gardner phase, there should be no significant barriers to a thorough experimentally testing of this phenomenon. In fact, an excess of low-energy excitations around jamming was reported in simulations~\cite{MKK08} in and experiments~\cite{KGMI10} of polydisperse crystals, before any proposal for a Gardner transition was even made. Revisiting these systems with a more fully developed Gardner physics in mind should therefore be reasonably productive. For instance, one could cyclically compress (without melting) crystals of agitated polydisperse, photoelastic disks and measure the change of the variability of the force network at jamming, depending on the decompression amplitude. Qualitatively different memory of previous force networks would be expected on either sides of the transition.

\section{RG for the Gardner criticality}
\label{sec:rg}
Gardner physics in mean-field theory is associated with the existence of soft, long-range excitations. One should thus expect that a full understanding of its criticality should be accessible from a renormalization group (RG) approach. Three different types of criticality are then to be considered: (i) the jamming transition, (ii) the Gardner phase, and (iii) the Gardner transition. Because  our understanding of the Gardner phase away from jamming is still fairly rudimentary, in this section we only consider types (i) and (iii).

\subsection{Jamming Criticality}
As we have seen in Sec.~\ref{sec:hs}, jamming criticality in finite dimensions is extremely well described by mean-field theory. In particular, all power-law scalings are essentially independent of spatial dimension for $d\ge 2$.  This superuniversal behavior is unusual. Most critical phenomena exhibit a strong dimensional dependence at low $d$.  A possible explanation for this  phenomenon has been suggested in Refs. \onlinecite{wyart2005rigidity,GLN12,HUZ19}. By comparing the fluctuations of the contact number $Z$ to its average -- as is commonly done in critical phenomena for the order parameter in a correlated volume -- one concludes that the upper critical dimension for jamming is $d_\mathrm{u}=2$.  Adding fluctuations on top of mean-field results and developing an RG treatment that would confirm this interpretation without input from numerical simulations remains, however,  an open challenge. A first-principle understanding of this remarkable phenomenon is therefore still lacking. Meanwhile, its lower critical dimension is still unexplored.

\subsection{Gardner Transition}

\begin{figure*}
\includegraphics[width=.95\textwidth]{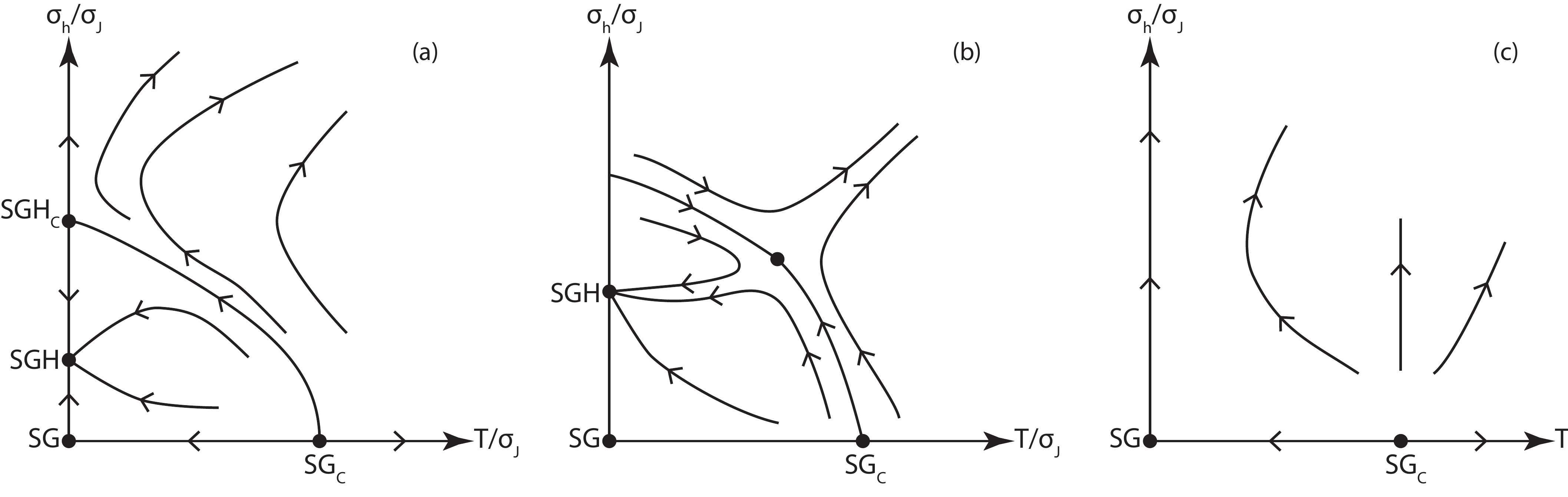}
\caption{Schematics of possible renormalization flow scenarios for the variances of the coupling constants, $\sigma_J$, and magnetic fields, $\sigma_h$, of a spin-glass model in $d<d_\mathrm{u}$. Note that even if the RG scheme starts with a uniform external field, a random field emerges after a few iterations\cite{angelini2015spin}. The points SG, SGH, $\mathrm{SG}_\mathrm{c}$, $\mathrm{SGH}_\mathrm{c}$ denote the FPs associated with the spin-glass phase, the spin-glass phase in a field, the critical point for spin glasses without field, and the critical point for spin glasses in a field, respectively.
(a) In this scenario, the transition in a field is present and the FP associated to it is a zero-temperature one, i.e. 
the couplings and the fields become much larger than the temperature under RG \cite{angelini2015spin}. (b) In this scenario, the transition in a field is present and the associated FP is standard, although non-perturbative \cite{charbonneau2017nontrivial}. (c) In this scenario, there is no FP in a field and, hence, no transition~\cite{moore2002p,yeo2012origin}. }
\label{fig:RG}
\end{figure*}

The Gardner transition in finite $d$ is under better theoretical control. Within their mean-field descriptions, the Gardner transition is analogous to the de Almeida-Thouless transition observed for spin glasses in an external magnetic field~\cite{dAT78}. Both transitions
display the same kind of order parameter and the same kind of collective behavior (or more precisely the same kind of instability of the disordered phase) and, hence, they share a same field theory of critical fluctuations~\cite{footnote1}.
This theoretical relationship and the activity on the Gardner transition of the last few years have triggered several new RG investigations of the problem. 

The modern starting point is the work of Moore et al.~\cite{moore2011disappearance} (see also Ref.~\onlinecite{urbani2015gardner}), which showed that the basin of attraction of the Gaussian fixed point (FP) associated with 
the mean-field description shrinks to zero when $d$ approaches the upper critical dimension, $d_\mathrm{u}=6$, from above. This results confirmed and detailed the original 1980 result of Bray and Roberts \cite{bray1980renormalisation} that the Gaussian FP disappears below $d_\mathrm{u}$,
and that no other perturbative FP emerges in $d<d_\mathrm{u}$. For many decades, this disappearance was interpreted as a signature of the non-existence of a critical transition in this context. However, that outcome is not the only possible one; a {\it non-perturbative} FP, and hence a critical transition, could instead replace it. This possibility has recently been investigated with different techniques: non-perturbative renormalization group \cite{charbonneau2017nontrivial}, high-order resummations \cite{charbonneau2018morphology}, and real space RG \cite{angelini2015spin,moore2002p,yeo2012origin}. Although the predictions of these different analyses differ quantitatively, they all show that a non-perturbative FP can actually exist in relatively low~$d$. 

These different RG scenarios are schematized in Fig.~\ref{fig:RG} for the flow parameters of a spin-glass model: the variance of the magnetic fields and of the magnetic couplings. (Although their counterparts for structural glasses give rise to an equivalent physics, they are somewhat less intuitive. They involve the difference in energy between different local particle arrangements, which leads to a generalized field and to elastic interactions related to these local strains \cite{fullerton}.)
The left and middle panels of Fig.~\ref{fig:RG} depict flows for non-perturbative FP with a zero-temperature \cite{angelini2015spin} and a finite-temperature\cite{charbonneau2017nontrivial} FP, respectively. The right panel depicts the flow in absence of a FP, and hence of a transition. The left and middle panels correspond to a case in 
which a Gardner transition can take place but has different properties: a zero-temperature FP implies activated dynamics 
scaling \cite{angelini2015spin} whereas a finite temperature one is associated to the usual power law relationship between time and length scales \cite{charbonneau2017nontrivial}. The right panel instead corresponds to a case without a bona-fide Gardner transition.   
An important issue to note is that even when a non-perturbative FP exists, the existence of a critical transition in the physical system at hand is not guaranteed. Its existence depends on whether the initial condition of the RG flow, i.e., the specific model considered, lies within the basin of attraction of the FP. As a result, 
although the properties of the critical transition are universal, its existence is not. For a given $d$, the transition could thus be 
present for amorphous solids but absent for certain spin-glass models, or present in finite-dimensional hard spheres but not in all soft sphere models. Probing the relationship between time and length-scales is a way to ascertain in experiments and in simulations to which basin of attraction, and to which scenario (a, b, c), a given model belong to. From a more theoretical point of view, one could try to map the model at hand into an effective theory, e.g. lattice spin models, that could then be studied more easily by RG and numerical simulations.  Neither has yet been reported.

The lower critical dimension of the transition, if present, is not known at this stage. There are, however,
results for the spin-glass transition without a field. In this case numerical simulations \cite{boettcher2004low}, real space renormalization group \cite{angelini2015spin}, and analysis of the low temperature phase \cite{astuti2018new} have converged  on the value $d_\mathrm{l}\simeq 2.5$. Assuming that adding an external field makes the transition more fragile -- a reasonable but not fully motivated proposal -- then one concludes that 
the lower critical dimension of the Gardner transition should be equal or higher than~$2.5$.  

In short, the current understanding is that a Gardner transition could take place in three dimensions and be related to a non-perturbative FP. Future investigations using more refined RG treatments, combined with new numerical simulations, will hopefully help close this long and intricate line of inquiry.  

\section{Gardner physics in Constraint Satisfaction Problems}
\label{sec:csp}

\begin{figure}
	\centering
	\includegraphics[width=\columnwidth]{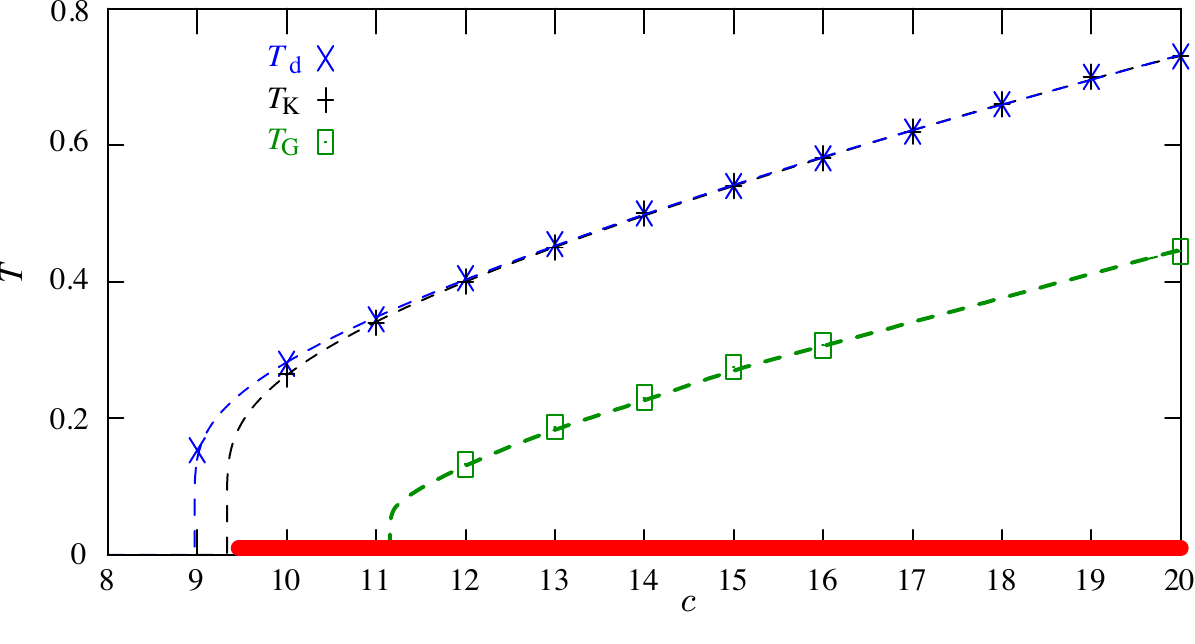}
	\caption{Sample phase diagram of the $q$-coloring CSP on a random regular graph with $q=4$, as a function of graph connectivity $c=2\alpha$ and temperature $T$.  The zero-temperature limit of this model illustrates the various types of instabilities discussed in the text. At low connectivity, $c<c_\mathrm{d}$ (and more generally above the line $T_{\rm d}(c)$), the space of solutions is ergodic. At $c_\mathrm{d}$, an ergodicity breaking transition gives rise to a clustered 1RSB phase. At $c_\mathrm{K}$ (and more generally at $T_{\rm K}(c)$), a Kauzmann transition gives rise to condensed clusters of solutions. Beyond the coloring-uncoloring (SAT-UNSAT) transition, $c_\mathrm{q}$, the zero-temperature energy becomes positive, hence not all constraints can typically be satisfied (red line, analogous to the jamming line in Fig.~\ref{fig:HSPD}). Finally, at $c_\mathrm{G}$ (and more generally at $T_{\rm G}(c)$ a Gardner transition gives rise to a Gardner phase, within which the typical states are marginally stable. Adapted from Ref.~\onlinecite{KZ08}.}
	\label{fig:PD}
\end{figure}

\begin{figure}
    \includegraphics[width=\columnwidth]{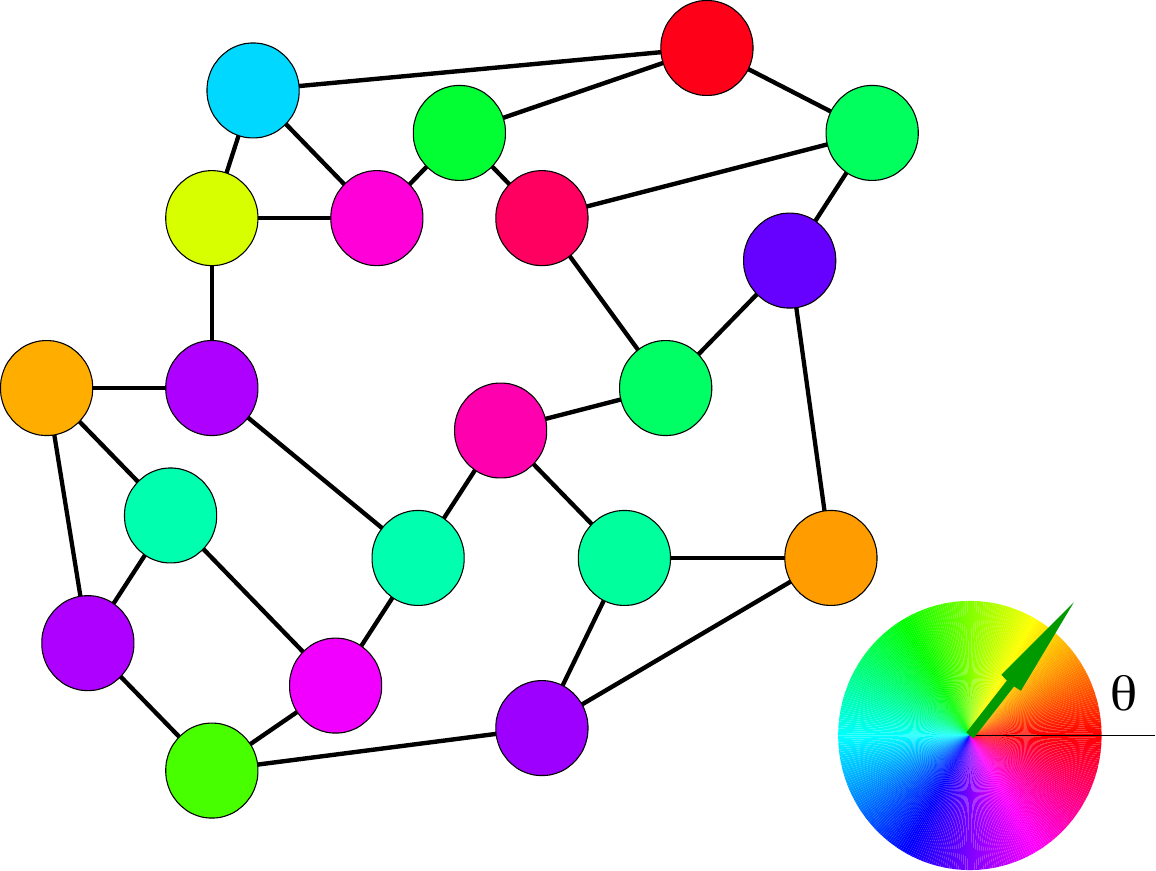}
    \caption{Illustration of the difference between continuous and discrete variables in the vertex coloring CSP. In the standard, discrete setting, $q$ colors are available to color the vertices of a given graph, such that no two adjacent vertices have the same color. (In the language of statistical physics, such a solution is the ground state of a Potts antiferromagnet.) In the continuous setting, colors are continuous variables on a circle and can therefore be defined by an angle $\theta$. The constraint is that the distance between colors on contiguous vertexes should be larger than a given threshold. For any finite graph, in the discrete problem the set of solutions is discrete and numerable. In the continuous setting this set is continuous, and its volume goes to zero at a jamming transition. Adapted from Ref.~\onlinecite{Y18}.
    }
    \label{fig:csp}
\end{figure}

In an interesting intellectual twist, the Gardner transition, which emerged from the study of spin glass models on fully-connected (all sites interacting with all others) graphs, reemerged in the study of spin glass models on diluted (i.e., locally tree-like) graphs, in parallel with its consideration in structural glasses.  
Theoretical interest in diluted spin glass models picked up speed in the
early 2000s, as interest for random versions of constraint satisfaction problems (CSP), to which they are intimately are related~\cite{MM09,handbook}, boomed. In the
most generic terms, CSP consist of finding assignments of variable that are subject to a set of
constraints.  Examples include graph-coloring, %K-SAT, 
constrained optimization, supervised learning, packing of hard objects etc.  CSP also provide the building blocks of computational complexity theory.~\cite{MM09,handbook} % K-SAT, for instance, was the first problem proved to be NP-complete~\cite{handbook}.

While complexity theory is only concerned  
with the \emph{worst} cases of a problem, interest in more representative \emph{typical} cases gave birth to a natural synergy between computer science and statistical physics. The tools developed for disordered systems were found to be especially useful to answer questions, such as:
If we draw a CSP randomly from an
ensemble of problems what is its typical complexity?
Does it admit a solution? What is the structure of the space of
solutions? 
This association emerges from the fact that random CSP problems can be
formulated as spin glass models with a Hamiltonian function being
zero for satisfying (SAT) assignments and quantifying constraint violation
for unsatisfying (UNSAT) ones.
The key parameter is then the ratio of 
the number of constraints to the number of variables, $\alpha$. Generically,
problems are SAT at small $\alpha$ and UNSAT at large $\alpha$, with
a sharp transition between the two regimes in the
thermodynamic limit of many constraints and variables~\cite{handbook,MM09}. As a further twist, the statistical physics of random CSP was also pioneered by Gardner, in her analysis of the single-layer neural network (perceptron)\cite{G88,DG88}. 

The set of SAT assignment of the spherical perceptron, if not empty, is convex and hence connected, glassy phases are not possible and
studying the SAT-UNSAT transition does not require RSB\cite{G88,DG88}.  For other prototypical CSP, such as %$k$-SAT and 
the coloring problem, however, 
the space of SAT assignments is not necessarily 
connected and only the implementation of 1RSB-based techniques\cite{MP01} enabled the exact computation of the SAT-UNSAT threshold
% of various %random K-SAT and 
%other problems
\cite{BMW00,MPZ02,KPW04}. 
In fact, many dilute discrete random CSP  belong to the class of
discontinuous (REM-like) spin glasses.  The route from the SAT to the UNSAT phase
as $\alpha$ increases 
is thus marked by several phase transitions, roughly following the RFOT scenario\cite{Kr07}: at low $\alpha$
the space of solutions (SAT assignments) is linked by small changes in variable assignments, i.e. ergodic; 
for  $\alpha>\alpha_\mathrm{d}$ this space first
undergoes a dynamical `clustering' transition at which solutions become organized in a large number of ergodic components but separated from one another by
changes to an extensive number of variable assignments; 
for $\alpha>\alpha_\mathrm{K}$ the space of solutions
is composed of a small number of clusters, and the SAT-UNSAT transition is reached at $\alpha>\alpha_\mathrm{q}$. 

The 1RSB ansatz used to analyze these models describes families of statistically identical 
states that lie at a same mutual distance. 
As in the case of spin glasses described in Sec.~\ref{sec:sg}, such a description can become unstable towards a more 
complex organization of states, both discussed in Ref.~\onlinecite{MPR04,KPW04}. 
In particular, a Gardner transition has been found in the space of solutions of a specific model of coloring of a random graph~\cite{KZ08}.

Using insights from the study of structural glasses\cite{FP16}, it was also realized that the SAT-UNSAT transition of CSP with continuous variables (cCSP) may also naturally display a critical scaling.  Because continuous variables make the space of
solutions also continuous, the volume of that space can indeed also smoothly shrink to zero at the SAT-UNSAT transition, as it does at the jamming transition (see, e.g., Fig.~\ref{fig:csp}). 
The analogy between cCSP and jamming was first noticed 
in the MK model of spheres\cite{MKK08} (discussed in Sec.~\ref{sec:hs}).
Although it was noted at the time
that low-frequency excitations around jammed
minima\cite{MKK08} are very similar to
those observed near jamming in more realistic models\cite{OLLN02,OSLN03,WNW05,WSNW05,BW09b}, this hint of a possible relation
between jamming criticality and the Gardner phase was overlooked.

It is rather in a \emph{non-convex} generalization of Gardner's perceptron problem that the relation between jamming and the 
Gardner phase in cCSP was first fully developed\cite{FP16}. 
Remarkably, isostatic jamming in this model is found in the Gardner phase and displays
the same critical properties as jamming in hard spheres. It has thus been conjectured that a unique universality class encapsulates all of critical jamming. 
A variety of models, including 
the high-dimensional vector spin models with excluded volume interaction
\cite{Y18},
feed-forward neural networks with a
single hidden layer and fixed hidden-to-output weights \cite{FHU18}, and deeper neural networks\cite{GSASBBW18} further support this claim. A more complete theory of critical jamming is, however, sorely needed. The study of jamming on diluted random cCSP, which has not yet been tackled, might also provide insights into the role of finite-dimensional effects. 

\section{Conclusion}
\label{sec:concl}
The last few years have seen an explosion of research motivated by the discovery of a Gardner transition in the solution of infinite-dimensional hard spheres. Although the initial hope of having a single transition unify the description of low-temperature glasses has been tempered, the insights it has provided are no less significant. As we have mentioned along this Perspective, a number of questions about the finite-dimensional Gardner transition remain actively pursued in model glass formers, in cCSP, and in RG. Of these, we are especially excited by the experimental tests made possible by recent theoretical and computational advances, and by the theoretical puzzle of the superuniversal jamming criticality. In addition, recent attempts at describing the jamming transition of slightly aspherical particles\cite{BIUWZ18}, which are surprisingly a lot more complex than simple spherical glass formers, suggest that the accompanying Gardner physics might be richer still. 

On the thirtieth anniversary of the untimely passing of Elizabeth Gardner, it is a sort of intellectual justice that her deep physical insights should still carry statistical mechanics forward.

\begin{acknowledgments}
Results discussed in this perspective article have been obtained within the Simons Collaboration ``Cracking the Glass Problem'', in a large collaborative research effort involving
several PIs other than the authors (J.~Kurchan, A.~Liu, S.~Nagel, G.~Parisi, and M.~Wyart) as well as many affiliates, postdocs and PhD students (A.~Altieri, C.~Brito, E.~DeGiuli, D.~Hexner, S.~Hwang, H.~Ikeda, J.~Kundu, E.~Lerner, Q.~Liao, M.~M\"uller, M.~Ozawa, T.~Rizzo, C.~Scalliet, B.~Seoane, J.~P.~Sethna, S.~Spigler, G.~Tsekenis, P.~Urbani, S.~Yaida, and H.~Yoshino).
We thank all of them as well as all the other colleagues with whom we have had the pleasure to  carry significant parts of the work described above,
in particular M.~C.~Angelini, K.~Daniels, O.~Dauchot, Y.~Jin, F.~Ladieu, and C.~Rainone.
We further acknowledge funding from the Simons Foundation (Grant \# 454933 LB; \#454935 GB \# 454937 PC; \# 454939 EC; \# 454941 SF; \# 454955 FZ), and from the European Research Council (ERC) under the European Union's Horizon 2020 research and innovation programme (grant agreement n. 723955 - GlassUniversality).
%%%%%KITP Acknowledgment%%%%%%
The writing of this publication was supported in part by the National Science Foundation under Grant No. NSF PHY-1748958.
\end{acknowledgments}

%\bibliography{./HS_SER,HSmerge,HS}

\begin{thebibliography}{166}
\bibitem [{foo({\natexlab{b}})}]{footnote1}%
  \BibitemOpen
  \href@noop {} {} \bibinfo {note} {We here consider the
  Gardner transition obtained by compressing or annealing an amorphous solid,
  not the transition that could take place for the ideal glass. The field
  theory of the latter differs slightly~\cite{urbani2015gardner}. Given that
  the main conclusions do not depend on this technicality, and that the former
  is in any event more experimentally relevant, we here neglect this subtle
  issue.}\BibitemShut {Stop}%
\end{thebibliography}

\end{document}